\documentclass{article}
\usepackage[utf8]{inputenc}
\usepackage{graphics,amsmath,amssymb,epsfig,epstopdf}

\title{Purely kinetic k-essence description of $c_s^2(w)$
barotropic fluid models}
\author{Dalibor Perkovi\'{c}$^{1,}$\thanks{dalibor.perkovic@zvu.hr}  and Hrvoje \v Stefan\v ci\'c$^{2,}$\thanks{hrvoje.stefancic@unicath.hr} }

\date{
\centering
$^{1}$ University of Applied Health Sciences, Mlinarska street 38, 10000 Zagreb, Croatia \\
\vspace{0.2cm}
$^{2}$ Catholic University of Croatia, Ilica 242, 10000 Zagreb, Croatia }

\begin{document}

\maketitle

\abstract{Purely kinetic k-essence models have been shown in the literature to be a field theory equivalent of barotropic fluid models of dark energy or dark matter-dark energy unification. In the modeling framework where the speed of sound squared of a barotropic fluid is modeled as a function of its Equation of State parameter, a systematic procedure of obtaining the Lagrangian density of an equivalent purely kinetic k-essence model is presented. As this modeling approach starts from the speed of sound, purely kinetic k-essence models can be constructed for which the speed of sound is in agreement with the observational constraints. Depending on the chosen functional form for the barotropic fluid speed of sound squared, analytically tractable examples of solutions for the purely kinetic k-essence Lagrangian density in parametric and closed form are obtained.}

\section{Introduction}

Available cosmological data strongly indicate that the expansion of the universe was accelerated in at least two phases: in the inflationary phase in the early universe \cite{Inflation1,Inflation2,Inflation3} and in the late-time phase of accelerated expansion \cite{SNIa1,SNIa2,CMB1,CMB2,BAO}. Identifying the mechanisms behind the accelerated cosmic expansion remains one of top priorities in physical cosmology. Numerous hypotheses on the nature of the said mechanism have been proposed and models of accelerated cosmic expansion elaborated. Two leading ideas for the mechanism of accelerated expansion are a component with a sufficiently negative pressure (called dark energy in the case of late-time accelerated expansion and inflaton field in the case of inflationary phase of accelerated expansion) \cite{DERev1,DERev2,DERev3,DERev4,DERev5} and the modified gravitational interaction \cite{ModGravRev1,ModGravRev2,ModGravRev3}. On the other hand, various cosmic phenomena, from galactic rotation curves to global expansion of the universe reveal the need for substantial amounts of additional matter (usually called dark matter) \cite{DM1,DM2,DM3,DM4}, modifications of gravitational interaction or even fundamental dynamic principles \cite{MOND1,MOND2,MOND3,MOND4}.   

An attractive idea is that the entire dark sector, i.e. that both dark energy and dark matter might be manifestations of the same component, sometimes called {\em dark matter-dark energy unification} or {\em unified dark matter} \cite{Bertacca}. Numerous interesting models in which the unified dark sector is described by a barotropic perfect fluid have been proposed. One of the most prominent ideas along this line of research is that of Chaplygin gas and its generalizations \cite{Chap1,Chap2,GenChap1,ModChap1,ModChap2,HybridChap,Umami}. For a general fluid based approach to the problem of unified dark sector see also \cite{LinderScherrer}. Despite the success of such models in describing the global evolution of the universe, many of these models exhibit considerable sound of speed at late time which significantly interferes  with the formation of structure at small cosmic scales \cite{GCGconstraint}. Some other interesting proposals of dark sector unification comprise superfluid dark matter \cite{superfluid} and dust of dark energy \cite{dustDE}. 

A prominent approach to dark sector description was developed in field theory models with scalar fields (for a review see e.g. \cite{Bertacca}). It was found that models with nonstandard kinetic terms called {\em k-essence} models, can be useful in the description of the inflationary phase \cite{CAP1,Garriga}, dark energy \cite{Chiba,CAP2,CAP3,Malquarti1,Malquarti2,Chimento,dePutter} and dark sector unification \cite{Scherrer,Chimento2,Armendariz,kessence2,Bertacca2,kessence3,Bertacca3}. In k-essence models, the Lagrangian density is in principle an arbitrary function of the scalar field $\phi$ and its kinetic term $X=\frac{1}{2} g^{\mu \nu} \nabla_{\mu} \phi \nabla_{\nu} \phi$, where $\nabla_{\mu}$ denotes covariant derivative, i.e. ${\cal L} = {\cal L}(X,\phi)$. Classes of general k-essence Lagrangian densities that have attracted most attention of researchers are ${\cal L} = V_1(\phi) F(X) - V_2(\phi)$ where dependences on $X$ and $\phi$ are factorized, and purely kinetic k-essence models where the k-essence Lagrangian density is a function of the kinetic term $X$ only, i.e. ${\cal L} = F(X)$. A special class of models motivated from string theories, so called tachyonic models \cite{tachyon1,tachyon2,tachyon3,tachyon4} can also be represented as k-essence models.  

A recently introduced and elaborated line of research approaches the problem of modeling barotropic fluid candidates for unified dark sector by specifying its speed of sound squared, $c_s^2$, as a function of its Equation of State (EoS) parameter, $w=p/\rho$, i.e. $c_s^2=c_s^2(w)$. Very initial studies \cite{Caplar} have shown that the problem of $c_s^2$ size and the accompanying acoustic horizon can be alleviated in the $c_s^2=\alpha (-w)^{\gamma}$ models which generalize the (generalized) Chaplygin gas model. In \cite{mi1}, the phenomenon of the cosmological constant barrier crossing was described in this formalism and in \cite{mi2} unified dark sector fluid models have been introduced with the speed of sound squared vanishing both in asymptotic past and in asymptotic future. Finally, in \cite{mi3} it was demonstrated that many of the dark energy parametrizations proposed in the literature are not compatible with dark energy being a barotropic fluid.   

In this work we combine advantages of k-essence and $c_s^2(w)$ approaches for the phenomenologically motivated modeling. K-essence models provide a microscopic formulation of the model in the framework of Lagrangian formalism of field theory. This fact simplifies comparison with other theoretically motivated or derived models and fosters establishing connections with fundamental physical theories. On the other hand, the $c_s^2(w)$ formalism starts from the modelling of the speed of sound as a function of $w$. For a known functional form of $c_s^2(w)$ one can determine the range of variation of the EoS parameter during cosmic expansion, even when the explicit analytic solution $w(a)$ is not available. Then, it is straightforward to obtain the range of variation of $c_s^2$ as the universe expands. Given these advantages, within the $c_s^2(w)$ formalism it is easy to construct models of unified dark sector which satisfy the observational constraints on the unified dark sector fluid speed of sound. In this paper, starting from the expressions introduced in \cite{dePutter}, we introduce a systematic analytic method of obtaining $F(X)$ of purely kinetic k-essence for a known $c_s^2(w)$ function. In this way, for a known $c_s^2(w)$ one can analytically determine the range of variation of $w$ and, consequently $c_s^2$, and thus guarantee that the speed of sound is in the allowed region. In the second step, the functions $X$ and $F$ can be obtained as parametric solutions with $w$ as a parameter and finally, depending on the invertibility of functions $X(w)$ and $F(w)$, implicit or explicit solutions for $F(X)$ can be obtained. 

The organization of the paper is the following. The first section contains the introduction and the outline of the main idea of the paper. The second section brings a brief overview of the general formalism of purely kinetic k-essence models. Then it elaborates the method of reconstruction of purely kinetic k-essence model for a known $c_s^2(w)$ function. The third section brings specific results for a number of concrete $c_s^2(w)$ models organized in two subsections: a subsection dedicated to parametric solutions and a subsection dedicated to closed form (implicit and explicit) solutions. In this subsection we also show that the studied models exemplify qualitatively different types of behavior of $c_s^2(a)$. The paper closes with the section Discussion and conclusions.   

\section{The reconstruction method}

\subsection{Purely kinetic k-essence models}


For purely kinetic k-essence models with the Lagrangian density ${\cal L}=F(X)$ one can find an equivalent perfect fluid representation \cite{Diez,Arroja,Ferreira}. In this representation the fluid four-velocity is $u_\mu = \frac{\nabla_\mu \phi}{\sqrt{2 X}}$, the energy density has the form
\begin{equation}
\label{eq:kessrho}
\rho=2 X F_{X}-F \, ,
\end{equation}
where the subscript $X$ stands for differentiation with respect to $X$ variable, whereas its pressure is
\begin{equation}
\label{eq:kessp}
p = F \, .
\end{equation}

The speed of sound in a purely kinetic k-essence models is 
\begin{equation}
\label{eq:kesscs2}
c_s^2=\frac{d p}{d \rho}=\frac{p_X}{\rho_X} = \frac{F_X}{F_X+2 X F_{XX}} \, ,
\end{equation}
and the parameter of the EoS is
\begin{equation}
\label{eq:kessw}
w=-\frac{F}{F-2 X F_X} \, .
\end{equation}

Inserting (\ref{eq:kessrho}) and (\ref{eq:kessp}) into the equation of continuity $d\rho + 3 (p+\rho) \frac{d a}{a} =0$ yields a relation \cite{Scherrer}
\begin{equation}
\label{eq:cont}
\frac{X}{X_0} \left(\frac{F_{X}}{F_{X,0}} \right)^2 \left(\frac{a}{a_0} \right)^6=1 \, .
\end{equation}

In the following subsection we elaborate an analytical approach more suitable to $c_s^2(w)$ models.  

\subsection{Representation of $c_s^2(w)$ models as purely kinetic k-essence models}


Starting from the definition of the sound of speed squared it is easy to obtain 
\begin{equation}
\label{eq:dpda}
a \frac{dp}{da}=-3(1+w) c_s^2 \rho \, ,
\end{equation}
which, using (\ref{eq:kessw}), leads to \cite{dePutter}
\begin{equation}
\label{eq:Fofa}
a \frac{d F}{d a} = -3 \frac{1+w}{w} c_s^2 F \, .
\end{equation}

On the other hand, the expression for the EoS parameter (\ref{eq:kessw}) gives
\begin{equation}
\label{eq:XFX}
X F_X = \frac{1+w}{2 w} p \, .
\end{equation}
Combining this expression with (\ref{eq:cont}) results in \cite{dePutter}
\begin{equation}
\label{eq:Xofa}
a \frac{d X}{d a} = - 6 c_s^2 X \, .
\end{equation}

The expressions (\ref{eq:Fofa}) and (\ref{eq:Xofa}) can be further transformed in forms better adapted for the approach in which the speed of sound squared is modelled as a function of the EoS parameter, $c_s^2=c_s^2(w)$. The equation of continuity, which can be written as 
\begin{equation}
\label{eq:wofa}
\frac{d w}{(1+w)(c_s^2-w)} = - 3 \frac{da}{a} \, ,
\end{equation}
allows elimination of the scale factor $a$ from (\ref{eq:Fofa}) and (\ref{eq:Xofa}) which then finally read
\begin{equation}
\label{eq:Fofw}
\frac{d F}{F} = \frac{c_s^2}{w(c_s^2-w)} d w \, ,
\end{equation}
\begin{equation}
\label{eq:Xofw}
\frac{d X}{X} = 2 \frac{c_s^2}{(1+w)(c_s^2-w)} d w \, .
\end{equation}

The expressions (\ref{eq:Fofw}) and (\ref{eq:Xofw}) represent the key novel extension of the approach presented in \cite{dePutter} which is especially suitable for the $c_s^2(w)$ models. For models defined by $c_s^2=c_s^2(w)$ these expressions, when their right sides are integrable, allow parametric definition of the purely kinetic k-essence models as $(F(w),X(w))$. If the expression for $X(w)$ is analytically invertible, the purely kinetic k-essence model can be obtained in an explicit form $F=F(X)$. If the expression for $F(w)$ is analytically invertible, the purely kinetic k-essence model can be obtained in an implicit form $X=X(F)$. In the next section we provide specific examples of $c_s^2(w)$ models for which analytic solutions for $F$ and $X$ functions can be obtained.  

The expressions (\ref{eq:Fofw}) and (\ref{eq:Xofw}) can be further expanded to give
\begin{equation}
\label{eq:Fofwexpanded}
\frac{d F}{F} = \left[ \frac{1}{c_s^2-w} + \frac{1}{w} \right]d w \, ,
\end{equation}
\begin{equation}
\label{eq:Xofwexpanded}
\frac{d X}{X} = 2 \left[ \frac{1}{c_s^2-w} +\frac{1}{1+w} -\frac{1}{(1+w)(c_s^2-w)} \right] d w \, .
\end{equation}

Combining (\ref{eq:Fofwexpanded}) and (\ref{eq:Xofwexpanded}) results in the equation
\begin{equation}
\label{eq:FX}
2 \frac{ d F}{F} - \frac{d X}{X} = 2 \frac{d w}{w} - 2 \frac{d w}{1+w} + 2 \frac{ d w }{(c_s^2-w)(1+w)} \, .
\end{equation}
Using (\ref{eq:wofa}) this equation can be further written as
\begin{equation}
\label{eq:FXa}
2 \frac{ d F}{F} - \frac{d X}{X} = 2 \frac{d w}{w} - 2 \frac{d w}{1+w} -6 \frac{ d a}{a} \, .
\end{equation}
Direct integration leads to the constraint (the subscript 0 hereafter refers to the present moment in the cosmic dynamics)
\begin{equation}
\label{eq:FXofa}
\left(\frac{F}{F_0} \right)^2 \frac{X_0}{X} = \left(\frac{w}{w_0} \right)^2 \left(\frac{1+w_0}{1+w} \right)^2 \left(\frac{a}{a_0} \right)^{-6} \, .
\end{equation}
For models in which $w$ can be obtained as a function of the scale factor, $w=w(a)$, Eq. (\ref{eq:FXofa}) provides $F^{2}/X$ as a function of $a$.

\section{Analytical solutions for $c_s^2(w)$ models }

In the next two subsections we present the application of the methodology introduced in the previous section to obtain solutions in parametric and closed form for concrete $c_s^2(w)$ models. In the considerations presented below, especially those that include inversion of expressions, multivalued functions denote all possible branches of these functions (e.g. $x^{1/2}$ stands for both $\sqrt{x}$ and $-\sqrt{x}$). The primary aim of such notation choice is to avoid too cumbersome expressions.


\subsection{Parametric solutions}

\subsubsection{The cosmological constant crossing model}

The model defined by
\begin{equation}
\label{eq:cs2CCcrossing}
 c_s^2=w + A (1+w)^B \, ,   
\end{equation}
was shown in \cite{mi1} to be able to describe the crossing of the cosmological constant barrier in certain parametric regimes. It should also be stressed that this model provides an effective description of the cosmological constant barrier crossing since its interpretation as a perfect fluid would lead to unphysical behavior of its speed of sound.

For parameter values $A \neq 0$, $B \neq 0$ and $B \neq 1$ the scale dependence of the EoS parameter is
\begin{equation}
\label{eq:wofaCCcrossing}
w=-1 + \left[(1+w_0)^{-B} + 3 A B \ln \frac{a}{a_0} \right]^{-\frac{1}{B}} \, .
\end{equation}
Inserting (\ref{eq:cs2CCcrossing}) into (\ref{eq:Fofw}) and (\ref{eq:Xofw}) leads to the specification of this model in the parametric form:
\begin{equation}
\label{eq:FofwCCcrossing}
F=F_0 \, \frac{w}{w_0} e^{\frac{1}{A(1-B)} [(1+w)^{-B+1} - (1+w_0)^{-B+1}]} \, ,
\end{equation}
\begin{equation}
\label{eq:XofwCCcrossing}
X=X_0 \, \left( \frac{1+w}{1+w_0}\right)^2 e^{\frac{2}{A(1-B)} [(1+w)^{-B+1} - (1+w_0)^{-B+1}]+\frac{2}{A B} [(1+w)^{-B} - (1+w_0)^{-B}]} \, .
\end{equation}

\subsubsection{CPL model}

We next apply our approach to a dark energy $w(a)$ parametrization introduced in \cite{Chevallier,Linder}, frequently referred to as the CPL model. The dependence of the EoS paremeter on the scale factor is given by  
\begin{equation}
\label{eq:CPL}
w(a)=w_0 + w_1 \left( 1-\frac{a}{a_0} \right) \, ,
\end{equation}
where $w_0$ and $w_1$ are parameters. This expression allows the calculation of $c_s^2(w)$ for this model which now reads
\begin{equation}
\label{eq:wofaCPL}
c_s^2=\frac{3 w^2 + 2 w+w_0+w_1}{3(1+w)} \, .
\end{equation}
Inserting (\ref{eq:wofaCPL}) into (\ref{eq:Fofw}) and (\ref{eq:Xofw}) yields solution for the $F(X)$ function in the parametric form: 
\begin{equation}
\label{eq:XofwCPL}
X=X_0 \, e^{-6(w-w_0)} \left(\frac{w_0+w_1-w}{w_1} \right)^{-6(w_0+w_1)} \left(\frac{1+w}{1+w_0} \right)^2 \, ,
\end{equation}
\begin{equation}
\label{eq:FofwCPL}
F=F_0 \, \frac{w}{w_0} \left(\frac{w_0+w_1-w}{w_1} \right)^{-3(1+w_0+w_1)} e^{-3(w-w_0)} \, .
\end{equation}
These solutions are consistent (up to the definition of constants) with the results obtained in \cite{Cardenas}.

\subsection{Closed form solutions}
 
\begin{figure}[!t]
\centering
\includegraphics[scale=0.45]{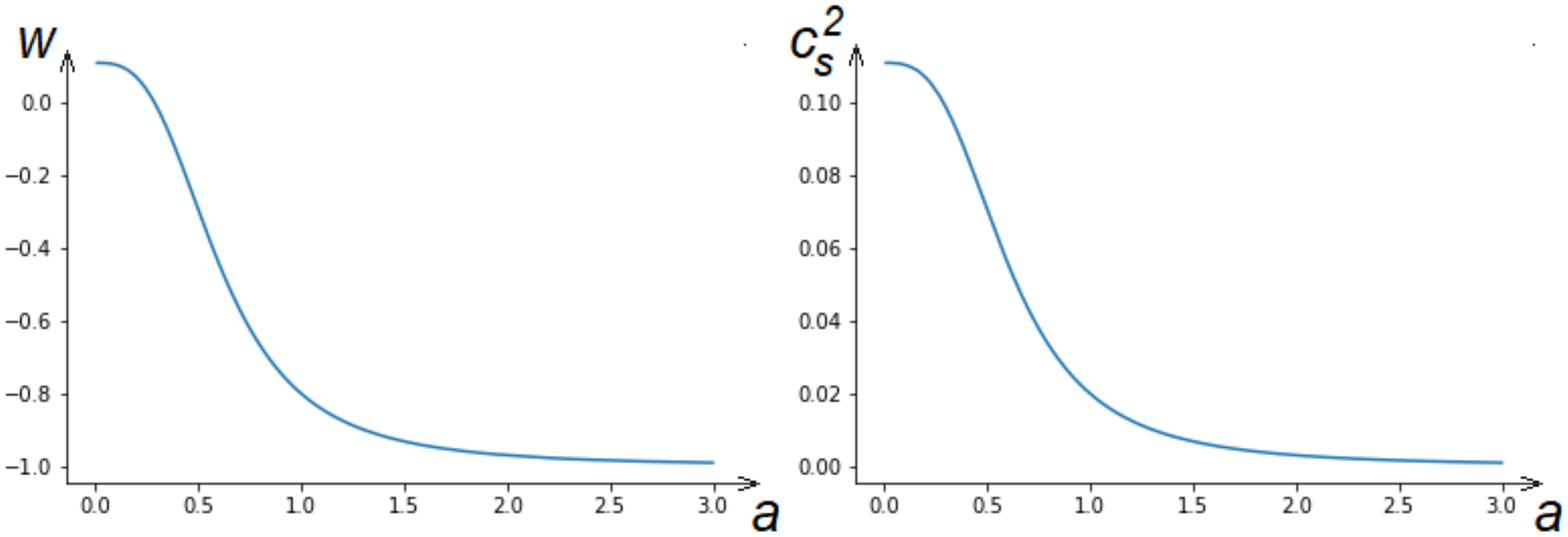}
\caption{The behavior of parameter of EoS and speed of sound squared for Model 1 with parameters $w_0=-0.8$ and $\alpha=0.1$.  
}
\label{fig_Model1}
\end{figure}

\begin{figure}[!t]
\centering
\includegraphics[scale=0.45]{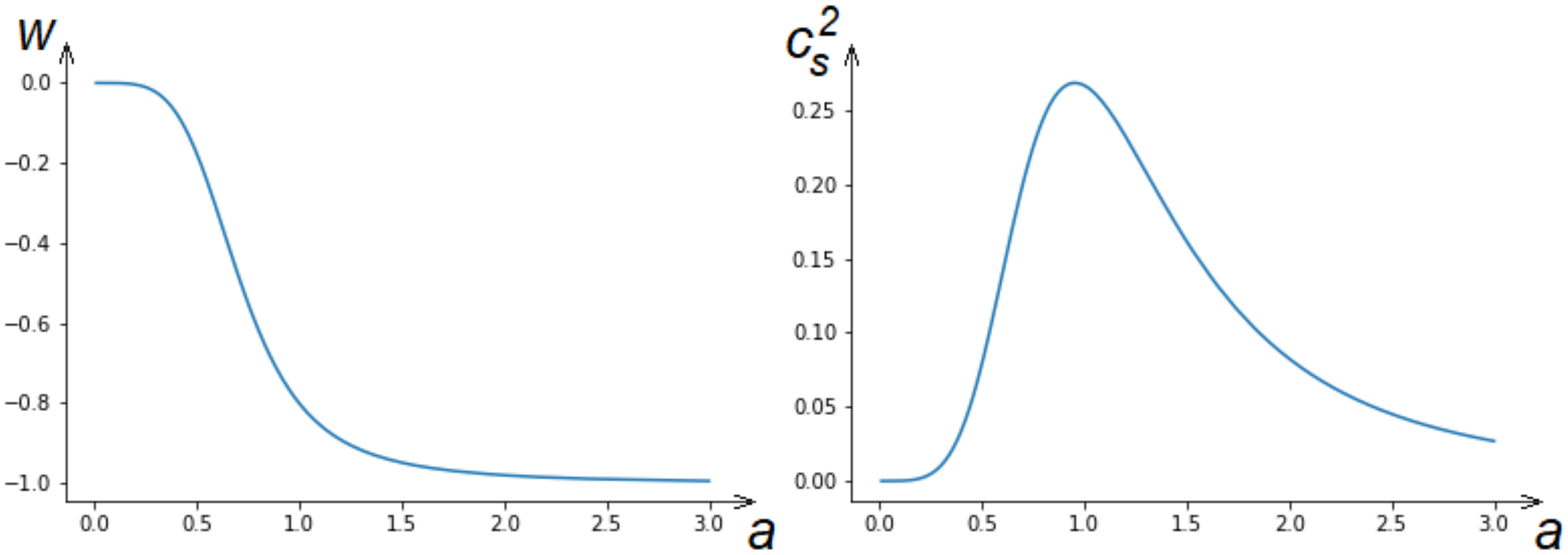}
\caption{The behavior of parameter of EoS and speed of sound squared for Model 2 with parameters $w_0=-0.8$, $\alpha=0.5$ and $w_*=-1.1$. 
}
\label{fig_Model2}
\end{figure}

\begin{figure}[!t]
\centering
\includegraphics[scale=0.45]{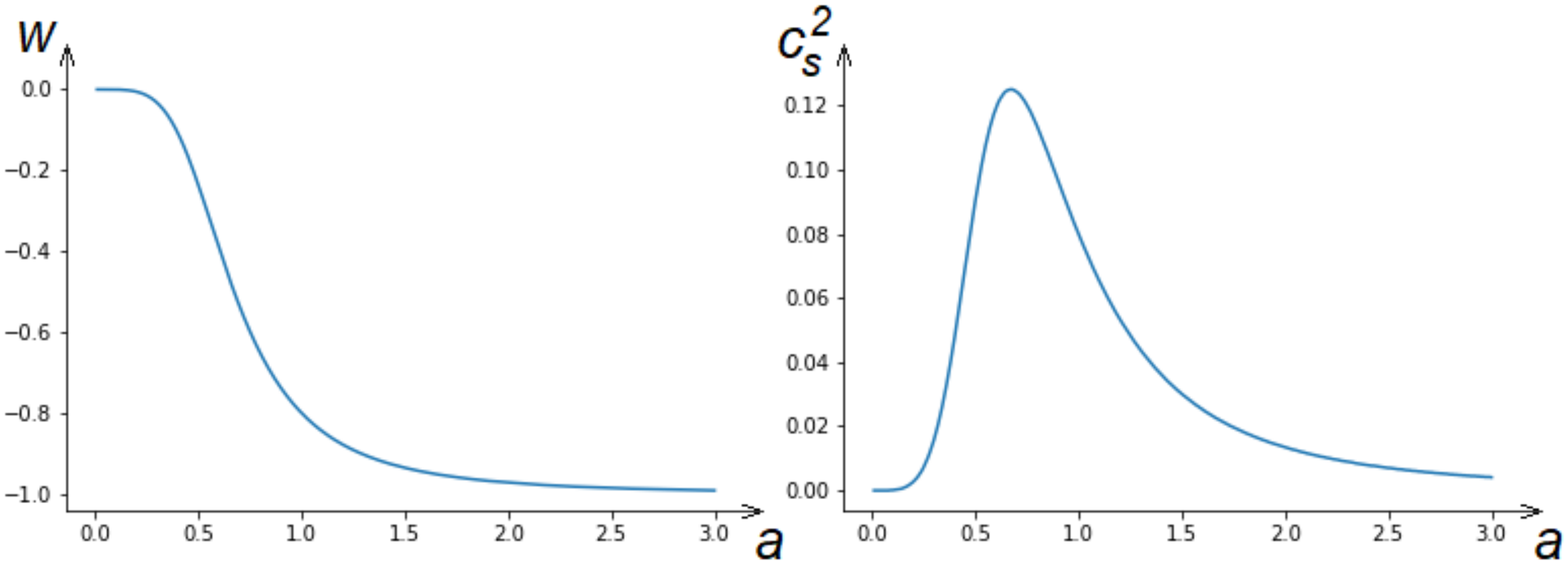}
\caption{The behavior of parameter of EoS and speed of sound squared for Model 3 with parameters $w_0=-0.8$ and $\alpha=-0.5$. 
}
\label{fig_Model3}
\end{figure}

\begin{figure}[!t]
\centering
\includegraphics[scale=0.45]{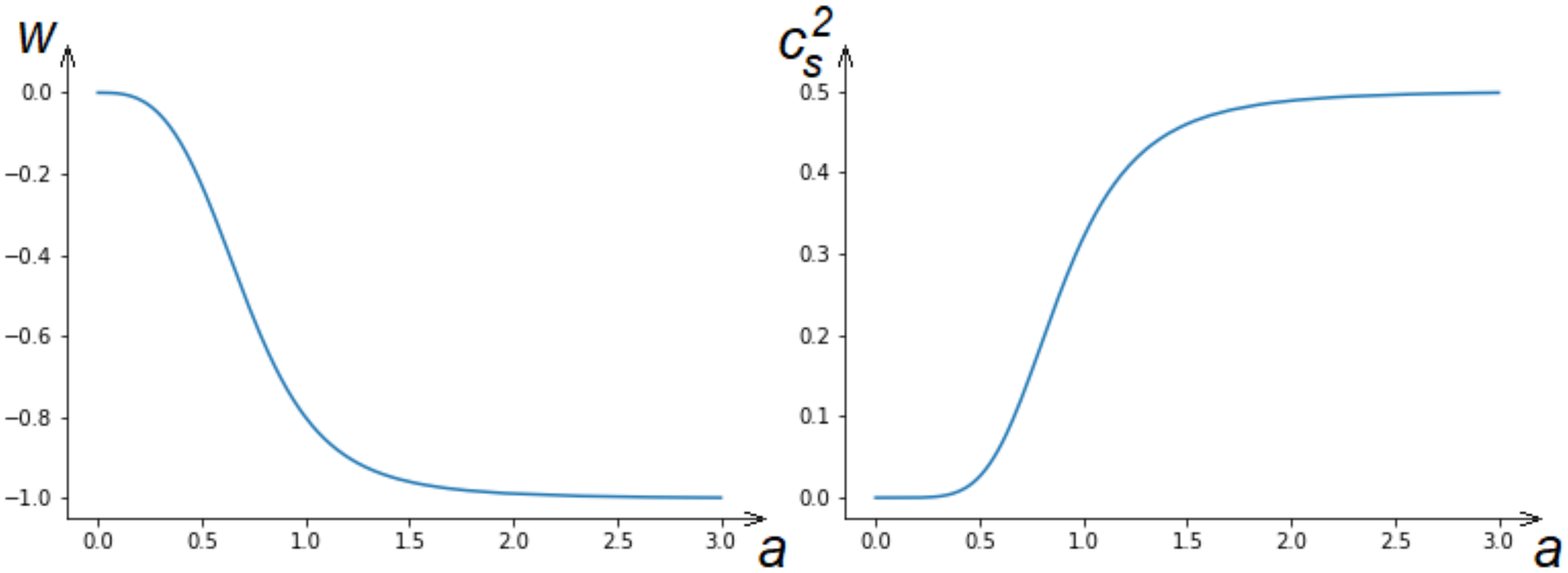}
\caption{The behavior of parameter of EoS and speed of sound squared for Model 4 with parameters $w_0=-0.8$, $\alpha=0.5$ and $\gamma=2$. 
}
\label{fig_Model4}
\end{figure}

In this subsection we present four models for which the closed form analytical solutions can be obtained. The primary aim of this subsection is to demonstrate $c_s^2(w)$ models for which the equivalent purely kinetic k-essence models can be analytically reconstructed. However, it is also important to stress that the presented models exemplify different types of dependence of $c_s^2(a)$. In Model 1, presented in Fig. \ref{fig_Model1}, the parameter of EoS interpolates between $\frac{\alpha}{1-\alpha}$ for $a \rightarrow 0$ and $-1$ for $a \rightarrow \infty$ and $c_s^2$ interpolates between $\frac{\alpha}{1-\alpha}$ for $a \rightarrow 0$ and $0$ for $a \rightarrow \infty$. In Models 2 and 3, presented in Figures \ref{fig_Model2} and \ref{fig_Model3}, the parameter of EoS interpolates between $0$ for $a \rightarrow 0$ and $-1$ for $a \rightarrow \infty$ whereas $c_s^2$ is asymptotically $0$, both for $a \rightarrow 0$ and $a \rightarrow \infty$. However, in these models $c_s^2$ has a maximum at some intermediate value of the scale factor. Finally, in Model 4, presented in Fig. \ref{fig_Model4}, the parameter of EoS interpolates between $0$ for $a \rightarrow 0$ and $-1$ for $a \rightarrow \infty$ and $c_s^2$ interpolates between $0$ for $a \rightarrow 0$ and $\alpha$ for $a \rightarrow \infty$.
Therefore, these models provide three qualitatively different types of $c_s^2(a)$ behavior and in the remainder of this subsection their analytical elaboration is presented.

 \subsubsection{Model 1}
 
 We first consider the model defined by
 \begin{equation}
\label{eq:cs2model1}
c_s^2=\alpha (1+w) \, .
\end{equation}
Let us first consider the general case $\alpha \neq 1$. Inserting (\ref{eq:cs2model1}) into (\ref{eq:wofa}) yields
\begin{equation}
\label{eq:wofa:model1}
w=-\frac{1-\frac{\alpha}{\alpha-1} \frac{1+w_0}{w_0+\frac{\alpha}{\alpha-1}} \left(\frac{a}{a_0} \right)^{-3}}{1-\frac{1+w_0}{w_0+\frac{\alpha}{\alpha-1}} \left(\frac{a}{a_0} \right)^{-3}} \, .
\end{equation}
The particular property of this model is that in the early universe the speed of sound differs from 0 (with the size of this difference controlled by the parameter $\alpha$) and in the late-time universe the speed of sound approaches 0 as $w$ approaches -1.

Following the procedure presented in preceding section, one can obtain the $X(w)$ function
\begin{equation}
\label{eq:Xofwmodel1}
X=X_0 \left(\frac{w+\frac{\alpha}{\alpha-1}}{w_0+\frac{\alpha}{\alpha-1}} \right)^{\frac{2 \alpha}{\alpha-1}} \, ,
\end{equation}
which can be readily inverted to give
\begin{equation}
\label{eq:wofXmodel1}
w= -\frac{\alpha}{\alpha-1} + \left(w_0+\frac{\alpha}{\alpha-1} \right) \left( \frac{X}{X_0}\right)^{\frac{\alpha-1}{2 \alpha}}\, .
\end{equation}
On the other hand, for $F(w)$ function one gets
\begin{equation}
\label{eq:Fofwmodel1}
F=F_0 \frac{w}{w_0} \left(\frac{w+\frac{\alpha}{\alpha-1}}{w_0+\frac{\alpha}{\alpha-1}} \right)^{\frac{1}{\alpha-1}} \, .
\end{equation}
Inserting (\ref{eq:wofXmodel1}) into (\ref{eq:Fofwmodel1}) finally yields
\begin{equation}
\label{eq:FofXmodel1}
F(X)=F_0 \left(\frac{X}{X_0} \right)^{\frac{1}{2 \alpha}} \left[-\frac{\alpha}{w_0(\alpha-1)} + \left(1+\frac{\alpha}{w_0(\alpha-1)}\right) \left(\frac{X}{X_0} \right)^{\frac{\alpha-1}{2 \alpha}}  \right] \, .
\end{equation}

For a particular parameter value $\alpha=1$ we obtain
\begin{equation}
\label{eq:wofamodel1alpha1}
w=-1 + (1+w_0) \left(\frac{a}{a_0} \right)^{-3} \, ,
\end{equation}
and
\begin{equation}
\label{eq:Xofwmodel1alpha1}
X=X_0 e^{2(w-w_0)} \, ,
\end{equation}
which after inversion gives us
\begin{equation}
\label{eq:wofXmodel1alpha1}
w=w_0 + \frac{1}{2} \ln \frac{X}{X_0} \, .
\end{equation}
For the function $F(w)$ we obtain
\begin{equation}
\label{eq:Fofwmodel1alpha1}
F=F_0 \frac{w}{w_0} e^{w-w_0} \, ,
\end{equation}
which finally results in 
\begin{equation}
\label{eq:FofXmodel1alpha1}
F(X)=F_0 \left( \frac{X}{X_0} \right)^{1/2} \left(1+ \frac{1}{2 w_0} \ln \frac{X}{X_0} \right) \, .
\end{equation}
 
 \subsubsection{Model 2}
 
 In this model the speed of sound squared as a function of $w$ is defined as  
\begin{equation}
\label{eq:cs2model2}
c_s^2=\alpha \frac{w(1+w)}{w_*-w} \, .
\end{equation}
For $w_* \neq -1$, $\alpha \neq -1$ and $\alpha \neq w_*$ the solution for the $w(a)$ function is
\begin{equation}
\label{eq:wofamodel2}
\left(\frac{w}{w_0} \right)^{\frac{w_*}{\alpha-w_*}} \left(\frac{w+w_1}{w_0+w_1} \right)^{-\frac{\alpha}{\alpha-w_*}} \frac{1+w}{1+w_0} = \left(\frac{a}{a_0} \right)^{-3} \, ,
\end{equation}
where  $w_1=\frac{\alpha- w_*}{1+\alpha}$.

Integration of (\ref{eq:Fofw}) gives 
\begin{equation}
\label{eq:Xofwmodel2}
X=X_0 \left(\frac{w+\frac{\alpha-w_*}{1+\alpha}}{w_0+\frac{\alpha-w_*}{1+\alpha}} \right)^{\frac{2 \alpha}{1+\alpha}}
\end{equation}
which can be inverted to provide an expression for $w$ as a function of $X$:
\begin{equation}
\label{eq:wofXmodel2}
w=-\frac{\alpha-w_*}{1+\alpha}+\left(w_0+\frac{\alpha-w_*}{1+\alpha} \right) \left( \frac{X}{X_0} \right)^{\frac{1+\alpha}{2 \alpha}} \, .
\end{equation}
Furthermore from (\ref{eq:Xofw}) it follows
\begin{equation}
\label{eq:Fofwmodel2}
F=F_0 \left( \frac{w}{w_0}\right)^{\frac{\alpha}{\alpha-w_*}} \left(\frac{w+\frac{\alpha-w_*}{1+\alpha}}{w_0+\frac{\alpha-w_*}{1+\alpha}} \right)^{-\frac{\alpha (1+w_*)}{(1+\alpha)(\alpha-w_*)}} \, ,
\end{equation}
which finally leads to the expression for $F(X)$:
\begin{equation}
\label{eq:FofXmodel2}
F(X)=F_0 \left( \frac{X}{X_0}\right)^{-\frac{1+w_*}{2(\alpha-w_*)}} \left[\frac{w_*-\alpha}{w_0 (1+\alpha)} +\left(1- \frac{w_*-\alpha}{w_0 (1+\alpha)}\right) \left(\frac{X}{X_0} \right)^{\frac{1+\alpha}{2 \alpha}} \right]^{\frac{\alpha}{\alpha-w_*}} \, .
\end{equation}

Next we present results for special values of parameters not covered by the results given above. For $w_*=-1$ the model is equivalent to $c_s^2=-\alpha w$ which corresponds to the generalized Chaplygin gas and we will not consider it any further. For $w_* \neq \alpha$ and $\alpha = -1$ we obtain the following results:
\begin{equation}
\label{eq:wofamodel2alpha-1}
\left(\frac{w}{w_0} \right)^{-\frac{w_*}{1+w_*}} \frac{1+w}{1+w_0} = \left(\frac{a}{a_0} \right)^{-3}\, ,
\end{equation}
\begin{equation}
\label{eq:Fofwmodel2alpha-1}
F=F_0 \left( \frac{w}{w_0} \right)^{\frac{1}{1+w_*}} e^{\frac{w-w_0}{1+w_*}} \, ,
\end{equation}
\begin{equation}
\label{eq:Xofwmodel2alpha-1}
X=X_0 e^{\frac{2}{1+w_*}(w-w_0)} \, .
\end{equation}
After inversion of the last expression we get 
\begin{equation}
\label{eq:wofXmodel2alpha-1}
w=w_0 + \frac{1+w_*}{2} \ln \frac{X}{X_0} \, ,
\end{equation}
which finally leads to 
\begin{equation}
\label{eq:FofXmodel2alpha-1}
F=F_0 \left( \frac{X}{X_0} \right)^{1/2} \left(1+\frac{1+w_*}{2 w_0} \ln \frac{X}{X_0} \right)^{\frac{1}{1+w_*}} \, .
\end{equation}

Finally we turn to the case $w_*=\alpha \neq -1$. The dependence of $w$ on the scale factor is given by the relation
\begin{equation}
\label{eq:wofamodel2alphawstar}
\frac{w_0}{w} \frac{1+w}{1+w_0} e^{-\frac{\alpha}{1+\alpha}(\frac{1}{w}-\frac{1}{w_0})} = \left(\frac{a}{a_0} \right)^{-3} \, .
\end{equation}
The dependence of $F$ and $X$ on $w$ are given by expressions
\begin{equation}
\label{eq:Fofwmodel2alphawstar}
F=F_0 \left(\frac{w}{w_0} \right)^{\frac{\alpha}{\alpha+1}} e^{\frac{\alpha}{\alpha+1}\left(-\frac{1}{w}+\frac{1}{w_0} \right)} \, ,
\end{equation}
\begin{equation}
\label{eq:Xofwmodel2alphawstar}
X=X_0 \left(\frac{w}{w_0} \right)^{\frac{2 \alpha}{1+\alpha}} \, .
\end{equation}
The inversion of the $X(w)$ expression yields 
\begin{equation}
\label{eq:wofXmodel2alphawstar}
w=w_0 \left(\frac{X}{X_0} \right)^{\frac{1+\alpha}{2 \alpha}} \, ,
\end{equation}
which finally leads to
\begin{equation}
\label{eq:FofXmodel2alphawstar}
F=F_0 \left(\frac{X}{X_0} \right)^{1/2} e^{\frac{\alpha}{(1+\alpha)w_0} \left(1-\left(\frac{X}{X_0} \right)^{-\frac{1+\alpha}{2\alpha}} \right)} \, .
\end{equation}

\subsubsection{Model 3}

The model defined by the following speed of sound squared
\begin{equation}
\label{eq:cs2model3}
c_s^2=\alpha w (1+w) \, ,
\end{equation}
will be considered for $\alpha \neq 1$ and $\alpha = 1$. 

For parameter values $\alpha \neq 1$ the $w(a)$ function is determined by the expression
\begin{equation}
\label{eq:wofamodel3}
\left(\frac{w}{w_0} \right)^{\frac{1}{\alpha-1}} \frac{1+w}{1+w_0} \left(\frac{w+1-\frac{1}{\alpha}}{w_0+1-\frac{1}{\alpha}} \right)^{-\frac{\alpha}{\alpha-1}} = \left(\frac{a}{a_0} \right)^{-3} \, .
\end{equation}
The expressions for $X(w)$ and $F(w)$ are
\begin{equation}
\label{eq:Xofwmodel3}
X=X_0 \left(\frac{w+1-\frac{1}{\alpha}}{w_0+1-\frac{1}{\alpha}} \right)^2 \, ,
\end{equation}
which can be inverted to give 
\begin{equation}
\label{eq:wofXmodel3}
w=-1+\frac{1}{\alpha} + \left(w_0+1-\frac{1}{\alpha} \right) \left(\frac{X}{X_0} \right)^{1/2}
\end{equation}
and
\begin{equation}
\label{eq:Fofwmdel3}
F=F_0 \left( \frac{w}{w_0} \right)^{\frac{\alpha}{\alpha-1}} \left(\frac{w+1-\frac{1}{\alpha}}{w_0+1-\frac{1}{\alpha}} \right)^{-\frac{1}{\alpha-1}} \, ,
\end{equation}
which leads to
\begin{equation}
\label{eq:FofXmodel3}
F=F_0 \left(\frac{X}{X_0} \right)^{-\frac{1}{2(\alpha-1)}} \left[-\frac{\alpha-1}{\alpha w_0}+ \left( 1+\frac{\alpha-1}{\alpha w_0} \right) \left(\frac{X}{X_0} \right)^{1/2} \right]^{\frac{\alpha}{\alpha-1}} \, .
\end{equation}

For the special case $\alpha=1$, the dependence of $w$ on $a$ is
\begin{equation}
\label{eq:wofamodel3alpha1}
\frac{w+1}{w} e^{-\frac{1}{w}} = \frac{w_0+1}{w_0} e^{-\frac{1}{w_0}}\left(\frac{a}{a_0} \right)^{-3} \, .
\end{equation}
The expressions for $F(w)$ and $X(w)$ are
\begin{equation}
\label{eq:Fofwmodel3alpha1}
F=F_0 \frac{w}{w_0} e^{-\frac{1}{w}+\frac{1}{w_0}} \, ,
\end{equation}
\begin{equation}
\label{eq:Xofwmodel3alpha1}
X=X_0 \left(\frac{w}{w_0} \right)^2 \, .
\end{equation}
After inversion of $X(w)$ which yields 
\begin{equation}
\label{eq:wofXmodel3alpha1}
w=w_0 \left( \frac{X}{X_0} \right)^{1/2} \, ,
\end{equation}
which finally results in
\begin{equation}
\label{eq:FofXmodel3alpha1}
F=F_0 \left( \frac{X}{X_0} \right)^{1/2} e^{\frac{1}{w_0} \left[1-\left(\frac{X}{X_0} \right)^{-1/2} \right]} \, .
\end{equation}

\subsubsection{Model 4}

Finally, we turn to the model considered in \cite{Caplar}:
\begin{equation}
\label{eq:cs2model4}
c_s^2=\alpha (-w)^\gamma \, .
\end{equation}

For $\gamma=1$ this model corresponds to generalized Chaplygin gas and model (\ref{eq:cs2model4}) contains the generalized Chalygin gas as a special case. Furthermore, for $\gamma=1$ and $\alpha=1$ the original Chaplygin gas model is recovered. 

We next turn our attention for $\gamma=2$. For $\alpha \neq -1$, the expression for the scaling of the EoS parameter $w$ is
\begin{equation}
\label{eq:wofagamma2model4}
\frac{w_0}{w} \left(\frac{w-\frac{1}{\alpha}}{w_0-\frac{1}{\alpha}} \right)^{\frac{\alpha}{\alpha+1}} \left( \frac{1+w}{1+w_0} \right)^{\frac{1}{\alpha+1}} = \left(\frac{a}{a_0} \right)^{-3} \, .
\end{equation}
The expression for $X(w)$ is
\begin{equation}
\label{eq:Xofwgamma2model4}
X=X_0 \left(\frac{w-\frac{1}{\alpha}}{w_0-\frac{1}{\alpha}} \right)^{\frac{2}{1+\alpha}} \left( \frac{w+1}{w_0+1}\right)^{\frac{2 \alpha}{1+\alpha}} \, ,
\end{equation}
whereas the expression for $F(w)$ is
\begin{equation}
\label{eq:Fofwgamma2model4}
F=F_0 \frac{w-\frac{1}{\alpha}}{w_0-\frac{1}{\alpha}} \, ,
\end{equation}
which after inversion leads to
\begin{equation}
\label{eq:wofFgamma2model4}
w=\frac{1}{\alpha}+\left(w_0-\frac{1}{\alpha} \right) \frac{F}{F_0} \, .
\end{equation}
This finally leads to an implicitly defined relation for $F(X)$:
\begin{equation}
\label{eq:XofFgamma2model4}
X=X_0 \left(\frac{F}{F_0} \right)^{\frac{2}{1+\alpha}} \left( \frac{1+\frac{1}{\alpha}+\left(w_0-\frac{1}{\alpha}\right) \frac{F}{F_0}}{1+w_0} \right)^{\frac{2 \alpha}{1+\alpha}} \, .
\end{equation}
For a special case $\alpha=1$ an explicitly defined $F(X)$ follows from (\ref{eq:XofFgamma2model4})
\begin{equation}
\label{eq:gamma3alpha1model4}  
\left( \frac{F}{F_0} \right)_{1,2} = \frac{-1 \pm \sqrt{1+(w_0^2-1)\frac{X}{X_0}}}{w_0-1}
\, .
\end{equation}

For $\gamma=2$ and $\alpha=-1$ the relation for $w(a)$ is 
\begin{equation}
\label{eq:wofagamma2alpha-1model4}
\frac{w}{1+w} e^{\frac{1}{1+w}} = \frac{w_0}{1+w_0} e^{\frac{1}{1+w_0}} \left(\frac{a}{a_0} \right)^3 \, .
\end{equation}
The expression for $F(w)$
\begin{equation}
\label{eq:Fofwgamma2alpha-1model4}
F=F_0 \frac{1+w}{1+w_0} \, ,
\end{equation}
which can be easily inverted to yield
\begin{equation}
\label{eq:wofFgamma2alpha-1model4}
w=-1+(1+w_0) \frac{F}{F_0} \, .
\end{equation}
The expression for $X(w)$ is
\begin{equation}
\label{eq:Xofwgamma2alpha-1model4}
X=X_0 \left(\frac{1+w}{1+w_0} \right)^2 e^{2\left(\frac{1}{1+w}-\frac{1}{1+w_0} \right)}\, .
\end{equation}
Combining (\ref{eq:wofFgamma2alpha-1model4}) and (\ref{eq:Xofwgamma2alpha-1model4}) finally gives expression for $F(X)$ in an implicit form:
\begin{equation}
\label{eq:XofFgamma2alpha-1model4}
X=X_0 \left(\frac{F}{F_0} \right)^2 e^{\frac{2}{1+w_0} \left(\frac{F_0}{F}-1 \right)}\, .
\end{equation}

Next we turn to the case $\gamma=\frac{1}{2}$. First we discuss the general case $\alpha \neq \pm 1$. The dependence of the EoS parameter on the scale factor is given by the relation
\begin{equation}
\label{eq:wofagamma1/2model4}
\left[ \frac{(-w)^{1/2}+\alpha}{(-w_0)^{1/2}+\alpha} \right]^{\frac{1}{1-\alpha^2}} \left[ \frac{(-w)^{1/2}+1}{(-w_0)^{1/2}+1} \right]^{\frac{1}{2(\alpha-1)}} \left[ \frac{1-(-w)^{1/2}}{1-(-w_0)^{1/2}} \right]^{-\frac{1}{2(1+\alpha)}} = \left(\frac{a}{a_0} \right)^{3/2} \, .
\end{equation}
The expression for the $X(w)$ function is
\begin{equation}
\label{eq:Xofwgamma1/2model4}
X = X_0 \left[ \frac{(-w)^{1/2}+\alpha}{(-w_0)^{1/2}+\alpha} \right]^{\frac{4 \alpha^2}{1-\alpha^2}} \left[ \frac{(-w)^{1/2}+1}{(-w_0)^{1/2}+1} \right]^{\frac{2 \alpha}{\alpha-1}} \left[ \frac{1-(-w)^{1/2}}{1-(-w_0)^{1/2}} \right]^{\frac{2 \alpha}{1+\alpha}} \, .
\end{equation}
The $F(w)$ function is 
\begin{equation}
\label{eq:Fofwgamma1/2model4}
F=F_0 \frac{w}{w_0} \left[ \frac{(-w_0)^{1/2}+\alpha}{(-w)^{1/2}+\alpha} \right]^2 \, ,
\end{equation}
which can be inverted to give
\begin{equation}
\label{eq:wofFgamma1/2model4}
w= -\frac{\alpha^2 \frac{F}{F_0}}{\left(1+\frac{\alpha}{(-w_0)^{1/2}} - \left(\frac{F}{F_0} \right)^{1/2} \right)^2 } \, .
\end{equation}
An implicitly defined $F(X)$ function can be obtained by inserting (\ref{eq:wofFgamma1/2model4}) into  (\ref{eq:Xofwgamma1/2model4}). Owing to length this expression is not explicitly written here. 

Next we discuss special values of $\alpha$. For $\alpha=1$ the $w(a)$ function becomes:
\begin{equation}
\label{eq:wofagamma1/2alpha1model4}
\left( \frac{(-w)^{1/2}-1}{(-w_0)^{1/2}-1} \right)^{1/2} \left( \frac{(-w)^{1/2}+1}{(-w_0)^{1/2}+1} \right)^{-1/2} e^{\frac{1}{(-w)^{1/2}+1} - \frac{1}{(-w_0)^{1/2}+1} } = \left(\frac{a}{a_0} \right)^{-3} \, .
\end{equation}
The $F(w)$ function is 
\begin{equation}
\label{eq:Fofwgamma1/2alpha1model4}
F=F_0 \frac{w}{w_0} \left[ \frac{(-w_0)^{1/2}+1}{(-w)^{1/2}+1} \right]^2 \, ,
\end{equation}
which can be inverted to give
\begin{equation}
\label{eq:wofFgamma1/2alpha1model4}
w= -\left[-1+\frac{1}{1-\frac{(-w_0)^{1/2}}{1+(-w_0)^{1/2}} \left(\frac{F}{F_0} \right)^{1/2}} \right]^2 \, .
\end{equation}
The expression for $X(w)$ is then
\begin{equation}
\label{eq:Xofwgamma1/2alpha1model4}
X = X_0 \frac{1-(-w)^{1/2}}{1-(-w_0)^{1/2}} \frac{1+(-w_0)^{1/2}}{1+(-w)^{1/2}} e^{-2 \left(\frac{1}{1+(-w)^{1/2}}-\frac{1}{1+(-w_0)^{1/2}}  \right)}\, .
\end{equation}
The implicitly defined $F(X)$ can be obtained by inserting (\ref{eq:wofFgamma1/2alpha1model4}) into  (\ref{eq:Xofwgamma1/2alpha1model4}).

Finally we turn to the case $\alpha=-1$. For the $w(a)$ function we obtain the relation
\begin{equation}
\label{eq:wofagamma1/2alpha-1model4}
\left(\frac{(-w)^{1/2}-1}{(-w_0)^{1/2}-1}\right)^{-1/2} \left(\frac{1+(-w)^{1/2}}{1+(-w_0)^{1/2}} \right)^{1/2} e^{- \left(\frac{1}{(-w)^{1/2}-1}-\frac{1}{(-w_0)^{1/2}-1} \right)}  = \left(\frac{a}{a_0} \right)^{-3}\, .
\end{equation}
The expression for $F(w)$ is 
\begin{equation}
\label{eq:Fofwgamma1/2alpha-1model4}
F=F_0 \frac{w}{w_0} \left[ \frac{(-w_0)^{1/2}-1}{(-w)^{1/2}-1} \right]^2 \, ,
\end{equation}
which after inversion gives
\begin{equation}
\label{eq:wofFgamma1/2alpha-1model4}
w= -\left[1+\frac{1}{-1+\frac{(-w_0)^{1/2}}{-1+(-w_0)^{1/2}} \left(\frac{F}{F_0} \right)^{1/2}} \right]^2 \, .
\end{equation}
The function $X(w)$ is
\begin{equation}
\label{eq:Xofwgamma1/2alpha-1model4}
X = X_0 \frac{(-w_0)^{1/2}-1}{(-w)^{1/2}-1} \frac{1+(-w)^{1/2}}{1+(-w_0)^{1/2}} e^{2 \left(\frac{1}{(-w)^{1/2}-1}-\frac{1}{(-w_0)^{1/2}-1}  \right)}\, .
\end{equation}
The implicitly defined $F(X)$ can be found by inserting (\ref{eq:wofFgamma1/2alpha-1model4}) into  (\ref{eq:Xofwgamma1/2alpha-1model4}).

\section{Discussion and conclusions}

The effectiveness of the approach presented in this paper depends crucially on the integrability of the right-hand expressions in Eqs (\ref{eq:Fofw}) and (\ref{eq:Xofw}). Additional insight into this problem can be obtained from the elaborated expressions for $d X/X$ and $d F/F$ given in (\ref{eq:Fofwexpanded}) and (\ref{eq:Xofwexpanded}) which consist of somewhat simplified expressions. Integrability can be achieved if functions $1/(c_s^2-w)$ and $1/((c_s^2-w)(1+w))$ are integrable.

Apart from the said advantages in analytical modeling of purely kinetic k-essence models, the approach defined by (\ref{eq:Fofw}) and (\ref{eq:Xofw}) can also be useful for numerical computation of the $F(X)$ function in the parametric form. Namely, $w$ is a convenient parameter since its variation during the entire cosmic expansion is limited to a finite interval (in many models $w$ is confined to the $[-1,0]$ interval).   

Another way of employing the described approach is to start by modelling $X(w)$ functions instead of $c_s^2(w)$ functions. This direction may lead to analytical solutions not accessible by other approaches. However, in this case the sufficiently small speed of sound cannot be guaranteed. Finally, an interesting question is to which extent the method presented in this paper can be applied to more general classes of k-essence models beyond the purely kinetic k-essence models.  

It should be further stressed that the main aim of this paper is to demonstrate the elaborated method of reconstruction of purely kinetic k-essence models for $c_s^2(w)$ models. To provide an analysis as complete as possible, all model parameter values were discussed although some of them may lead to models with nonphysical properties (such as negative speed of sound squared). These models have to be separately tested against the empirical data to determine the allowed parametric region.     

In conclusion, we demonstrated how the formalism of modeling speed of sound squared of dark matter-dark energy models as a function of its EoS parameter can be useful in obtaining purely kinetic k-essence models with an observationally acceptable speed of sound. In several examples we have shown how specific choices of $c_s^2(w)$ functions can be analytically translated into $F(X)$ functions of purely kinetic k-essence models. Depending on the functional form of $c_s^2(w)$, the $F(X)$ function can be obtained in the parametric or closed form. This approach integrates the theoretical advantages of field theoretic approaches such as k-essence models and phenomenological control over the speed of sound in the $c_s^2(w)$ formalism and its compatibility with the current observational constraints from the cosmic large scale structure.

\end{document}